\documentclass[amssymb,amsmath,aps,11pt,tightenlines,longbibliography,notitlepage]{revtex4-1}
\usepackage{graphicx}

\begin{document}

\title{Observation of a transition from a topologically ordered \\
to a spontaneously broken symmetry phase}

\renewcommand{\thefootnote}{\fnsymbol{footnote}}

\author{N. Samkharadze$^{1,}$\footnotemark[1], 
K.A. Schreiber$^{1,}$\footnote[1]{These authors contributed equally.} 
G.C. Gardner$^{2,3}$, M.J. Manfra$^{1,2,3,4}$, E. Fradkin$^5$, and 
G.A. Cs\'{a}thy$^{1,3}$}

\affiliation{
${}^1$Department of Physics and Astronomy, Purdue University, West Lafayette, IN 47907, USA \\
${}^2$School of Materials Engineering, Purdue University, West Lafayette, IN 47907, USA \\
${}^3$Birck Nanotechnology Center Purdue University, West Lafayette, IN 47907, USA \\
${}^4$School of Electrical and Computer Engineering, Purdue University, West Lafayette, IN 47907, USA \\
${}^5$Department of Physics and Institute for Condensed Matter Theory, University of Illinois, Urbana, IL 61801-3080, USA 
}

\date{\today}

\begin{abstract}
Until the late 1980s, phases of matter were understood in terms of Landau's symmetry breaking theory. 
Following the discovery of the quantum Hall effect the introduction of a second class of phases, those 
with topological order, was necessary. Phase transitions within the first class of phases involve a change in symmetry, 
whereas those between topological phases require a change in topological order. However, in rare cases transitions 
may occur between the two classes 
in which the vanishing of the topological order is accompanied by the emergence of a broken symmetry. 
Here, we report the existence of such a transition in the two-dimensional electron gas. When tuned by
hydrostatic pressure, the $\nu=5/2$ fractional quantum Hall state, believed to be a prototype 
non-Abelian topological phase, gives way to a quantum Hall nematic phase. 
Remarkably, this nematic phase develops spontaneously, i.e. in the absence of any externally applied symmetry breaking fields.
\end{abstract}

\maketitle

Numerous condensed matter systems exhibit phases with 
broken rotational symmetry, often referred to as electronic nematic phases \cite{fogler,moessner,kivelson,fradkin,lilly,du}. 
These phases may have additional broken symmetries \cite{fradkin}. The intertwining of
nematic order and other orders has recently elicited heightened interest. For example,
nematic order and superconducting order accompany each other in
high-$T_C$ superconductors \cite{high1,high3,high4} and may coexist in confined $p$-wave superfluids \cite{he3}.

The traditional phases of matter, such as the nematic phase or superconductors, are  distinguished by their
order parameters \cite{landau}. However, certain phases cannot be described by a local order parameter.
These phases, referred to as topological phases, are instead classified according to their topological order as measured 
by topological invariants \cite{wen1}.
Topological phases often have a degenerate ground state, a gap in their excitation spectrum, and possess current
carrying edge states. The integer \cite{klitzing}
and fractional quantum Hall states (FQHS) \cite{tsui} forming in the two-dimensional electron gas (2DEG) are prototypical topological phases.

A phase transition from a topological phase to a nematic phase
is expected to violate Landau's rules and may be of a novel kind since both the symmetry and 
the topological order need to change across such a transition. 
The FQHS at $\nu=5/2$, believed to be a special topological phase with non-Abelian properties
\cite{read,willett,panq}, is expected to have an instability toward a nematic phase \cite{haldane2000,wan,wang}.
In transport the broken rotational symmetry of a nematic phase is manifest in an anisotropic resistance.
A transition at $\nu=5/2$ from the isotropic FQHS to an anisotropic state
so far has only been observed
in the presence of a symmetry breaking in-plane magnetic field
\cite{pan99,lilly99,friess14,pan15,xia10,xia11,liu13}.
However, an externally applied symmetry breaking field is always expected to favor
the associated broken symmetry phase.
An intriguing question is, therefore, whether or not a transition from a topological phase to a 
broken symmetry phase may occur spontaneously, i.e. {\it in the absence} of any symmetry breaking fields. 

We investigate possible instabilities of the FQHS at $\nu=5/2$ by magnetoresistance measurements
under hydrostatic pressure $P$. When a 2DEG of density $n$ is exposed
to a perpendicularly applied magnetic field $B_{\perp}$, a number $\nu=nh/eB_{\perp}$ of
equidistant Landau levels will be filled \cite{qh}. We focus on the region called the second
Landau level, specifically the range of Landau level filling factors $2 < \nu < 3$.
The filling factor of interest $\nu=5/2$ is located in the middle of this range.
In Fig.1 we show the longitudinal resistance measured at three different pressures
at a temperature $T \simeq 12$~mK.
The longitudinal resistance is monitored along two mutually perpendicular crystallographic directions of the GaAs: 
$R_{XX}$ is obtained with the current bias $I$ applied and voltage drop measured 
along the [1$\bar{1}$0] crystal direction, whereas $R_{YY}$ is measured along the [110] direction. 
Details of our sample and the pressure cell used can be found in Methods.

In Fig.1a we show the magnetoresistances at $P=6.95$~kbar. The longitudinal resistances $R_{XX}$ and $R_{YY}$
exhibit sharp minima at $\nu=5/2$. As shown in the Supplement, these resistance minima are accompanied by a
plateau at $2h/5e^2$ in the Hall resistance, indicating therefore a fractional quantum Hall ground state at $\nu=5/2$ \cite{willett,panq}.
We note that $R_{XX}$ and $R_{YY}$ measured in the vicinity of $\nu=5/2$ along the different 
sample edges are nearly equal. We thus find that, similarly to measurements performed on samples in the 
ambient \cite{lilly,du}, the longitudinal resistance near $\nu=5/2$ measured at $P=6.95$~kbar is isotropic and 
therefore the ground state is a rotationally invariant FQHS. 

\begin{figure}
\def\ffile{Fig1.pdf}
\includegraphics[width=1\textwidth]{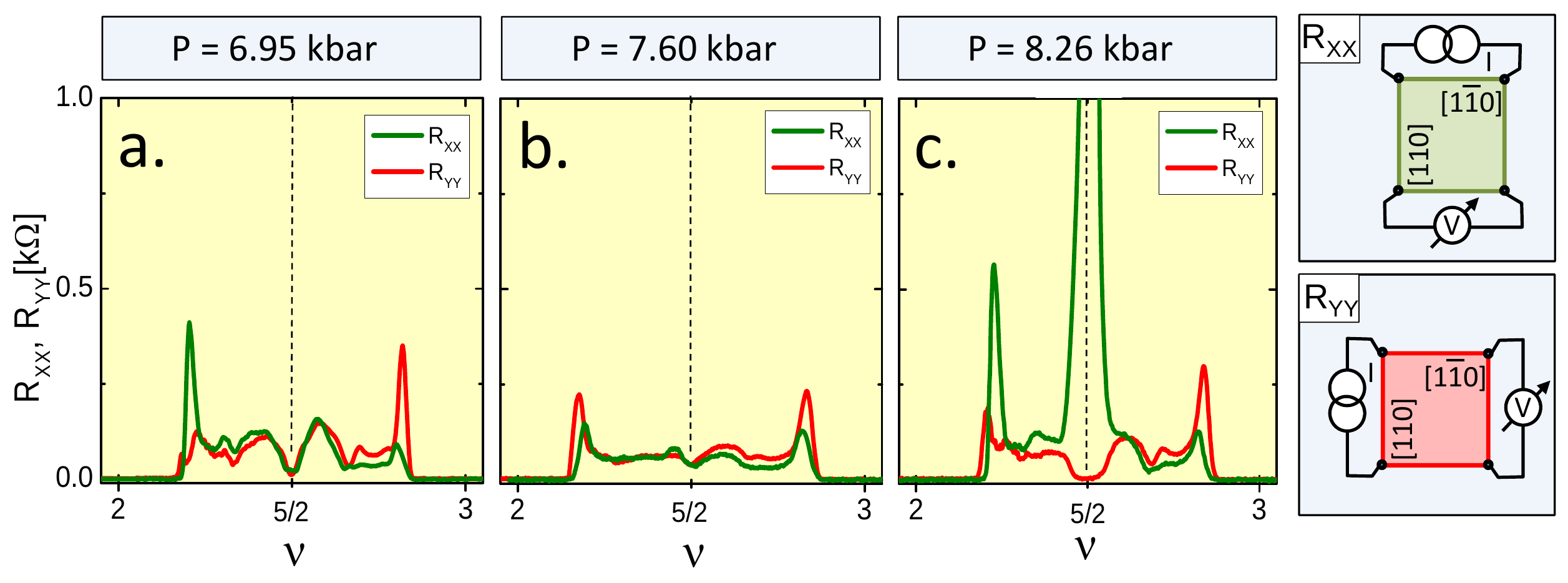}
\caption{Dependence of the magnetoresistance on hydrostatic pressure $P$ in the second Landau level.
The green traces show $R_{XX}$, while the red traces show $R_{YY}$
as measured along two mutually perpendicular crystallographic directions of GaAs.
Diagrams in the right indicate the circuit configurations used to obtain $R_{XX}$ and $R_{YY}$.
As the pressure is increased, at $\nu=5/2$ we observe the following sequence of ground states:
an isotropic FQHS, a nearly isotropic Fermi liquid, and the nematic phase.
The data is taken at $T \simeq 12$~mK.}
\label{\ffile}
\end{figure}

As the pressure is increased to $P=7.60$~kbar, the longitudinal resistance near $\nu=5/2$ remains isotropic.
However, as seen in Fig.1b, the strong minima in $R_{XX}$ and $R_{YY}$ are no longer present at this pressure.
The finite and isotropic resistance at $\nu=5/2$ exhibiting a weak dependence on the filling factor
is indicative of proximity of the ground state to a compressible isotropic Fermi liquid. 
Our data at $\nu=5/2$ shown in Fig.2b suggests therefore that the fractional quantum Hall 
ground state at $\nu=5/2$ approaches an instability.

A further increase in pressure to $P=8.26$~kbar causes a strong minimum to
reappear in $R_{YY}$ at $\nu=5/2$. As seen in Fig.1c, this minimum in $R_{YY}$ is visibly wider, i.e.
spans a larger range of filling factors, than that at $P=6.95$~kbar. The most dramatic change, 
however, is in $R_{XX}$, which in contrast to lower pressure data, exhibits a pronounced peak at $\nu=5/2$. 
The anisotropic resistance observed at $\nu=5/2$, characterized by an extremely large
ratio $R_{XX}/R_{YY}=1150$, signals the onset of a ground state which breaks rotational symmetry. 
The evolution of the magnetoresistance at $\nu=5/2$ shown in Fig.1 is therefore suggestive of a phase transition 
from a rotationally invariant FQHS, most likely a non-Abelian topological phase \cite{read,willett,panq}, to an 
anisotropic phase which breaks rotational symmetry.                
 
In 2DEGs with half filled Landau levels we differentiate between two types of anisotropies: 
spontaneous and induced anisotropy. The ground states of the 2DEG associated with these anisotropies
bear a close resemblance to the spontaneous and induced magnetism in an interacting spin system.
In the absence of an externally applied magnetic field, the spin system exhibits
spontaneous symmetry breaking which manifests in a sharp phase transition
between the ordered ferromagnet and the disordered paramagnet. 
In contrast, the development of the ordered phase with the application of an external magnetic field
is not associated with a thermodynamic singularity. 
In the 2DEG spontaneous anisotropy develops 
in the absence of any externally applied symmetry breaking fields
at $\nu=9/2, 11/2, 13/2, 15/2, ...$ \cite{lilly,du,heterostructure} and also at $\nu=7/2$ \cite{pan14}
at low enough temperatures.
The ground state at these filling factors is the well-known quantum Hall nematic phase, also referred to in the literature as the stripe phase \cite{fogler,moessner}.
Contrary to the behavior at the above filling factors, the ground state at $\nu=5/2$ was always found to be isotropic 
 in the absence of a symmetry breaking field \cite{lilly,du}.
Induced anisotropy at $\nu=5/2$ appears, however, with the application of an external symmetry breaking field.
An in-plane magnetic field at $\nu=5/2$ induces a compressible nematic-like anisotropic phase
\cite{pan99,lilly99,friess14,pan15} or an incompressible nematic FQHS \cite{xia10,xia11,liu13}.
Anisotropy also appears with the application of uniaxial strain, another symmetry breaking field \cite{leo11}.

In contrast to these prior experimental results, the anisotropy we observe at $\nu=5/2$
in Fig.1c has clearly developed spontaneously.
Indeed, because of  the hydrostatic nature of the applied pressure, in our experiment the rotational symmetry 
in the plane of the 2DEG is not broken by any external fields.
An unintentional in-plane magnetic field may appear in our experiment if our sample
tilts inside our cell during the compression process generating the high pressures. However,
the isotropic resistance near $\nu=5/2$ at $P=6.95$ and 7.60~kbar attests that this is not the case.
We therefore report a pressure-tuned spontaneous transition at $\nu=5/2$ from an isotropic FQHS to a quantum Hall nematic phase through an isotropic Fermi liquid phase. Since at $T=12$~mK the isotropic liquid is observed in an extremely narrow range of pressures, our data is suggestive of a direct quantum phase transition from the FQHS to the nematic phase in the limit of zero temperatures.

From the point of view of the symmetry, both the FQHS-to-nematic and the paramagnet-to-ferromagnet 
transitions occur spontaneously.
However, a notable difference between these two transitions is that in our data 
the collapse of the ordered nematic phase is accompanied by the emergence of
a topologically ordered phase rather than a disordered isotropic phase.
The FQHS-to-nematic transition we observe at $\nu=5/2$ is thus an example of a phase transition which involves
the change of both the topological as well as the rotational order across the transition. Such a phase transition was predicted in Ref.\cite{haldane2000}. 
However, because of the boundary conditions used
in this numerical work which break the rotational symmetry, the stripe phase could not be distinguished from the nematic phase and the phase transition could not be characterized. Our observations are incompatible with a direct first order phase transition from the FQHS to the nematic phase, but are compatible with either a direct continuous transition between these two phases or with an intercalation of an isotropic Fermi liquid between these two phases. In the former case we think that 
the quantum critical point is necessarily described by an exotic theory (not based on the Landau picture)
due to the interplay of the nematic order and the emergent topological order in the non-Abelian FQHS. We note that similar exotic transitions have been proposed in topologically ordered states \cite{ran} and in a generalized quantum dimer model \cite{fr2}.

The difference between the spontaneous and induced anisotropic phases at half filled Landau levels 
is further highlighted by their contrasting magnetotransport signatures. While both manifest in
anisotropic magnetoresistance,
a peculiarity of the spontaneous anisotropy is that it develops 
over a limited span of filling factors $\Delta \nu \simeq 0.15$ centered on a half integer filling factor \cite{lilly,du}.
In contrast, the resistance anisotropy induced by  an external in-plane magnetic field at $\nu=5/2$
is present over a considerably wider range of filling factors
$\Delta \nu \simeq 0.6$ \cite{pan99,lilly99,friess14,pan15,xia10,xia11,liu13}.
The observed anisotropy at $P=8.26$~kbar shown in Fig.1c occurring
over a narrow range of filling factors $\Delta \nu \simeq 0.15$ 
is consistent with our earlier conclusion
that the ground state at $\nu=5/2$  is a genuine quantum Hall nematic phase \cite{fogler,moessner}, 
similar to that observed at $\nu=9/2$ \cite{lilly,du}.

\begin{figure}
\def\ffile{Fig2.pdf}
\includegraphics[width=1\textwidth]{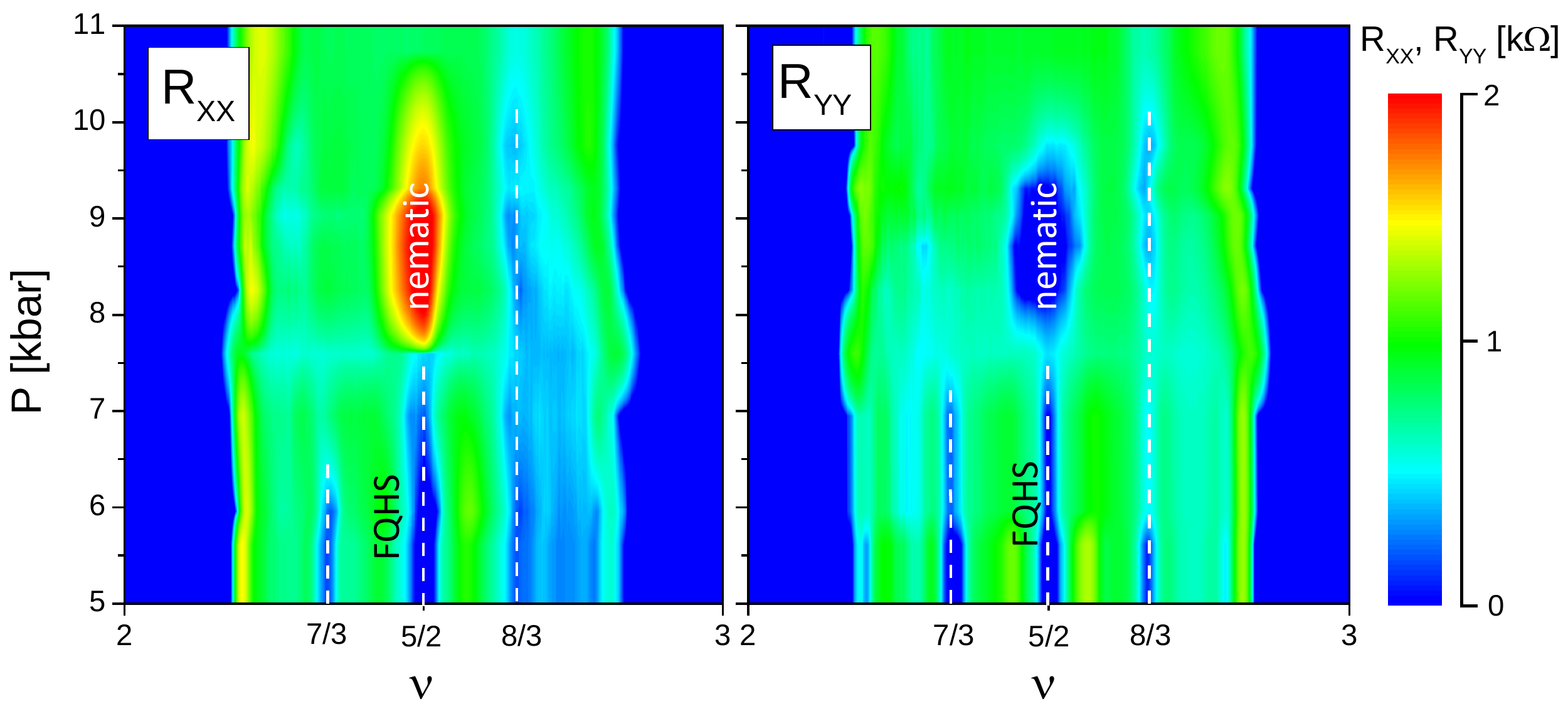}
\caption{Contour plots of the magnetoresistance against the pressure and filling factor at  $T \simeq 12$~mK.
At $\nu=5/2$ we observe a rotationally invariant FQHS
at $P<7.8$~kbar, the nematic phase at $7.8$~kbar$<P<10$~kbar, and an isotropic
Fermi liquid at $P>10$~kbar. The nematic phase develops in a narrow range of filling factors $\Delta \nu \simeq 0.15$
centered around $\nu=5/2$.}
\label{\ffile}
\end{figure}

The evolution of the two longitudinal resistances $R_{XX}$ and $R_{YY}$ is captured over a larger pressure range
in the contour plot Fig.2. We focus on the behavior along the line at $\nu=5/2$. At low pressures  
the FQHS is shown as a narrow vertical blue line. 
As the pressure is increased, the $\nu=5/2$ FQHS weakens and past a critical pressure
estimated to be $P_C \simeq 7.8 \pm 0.2$~kbar the nematic phase is stabilized.
The nematic phase is seen in Fig.2 as a red island in $R_{XX}$ and as a blue basin in $R_{YY}$. 
Our data at $\nu=5/2$ shown in Fig.2 reinforces the possibility of
a direct quantum phase transition from a FQHS to the nematic phase 
as the pressure is tuned through its critical value $P=P_C$.

In Fig.2 the region of stability for the nematic phase is centered near $P \simeq 8.7$~kbar.
The nematic phase is weakened by a further increase in pressure until it
disappears at an extrapolated value of $P \simeq 10$~kbar. 
Past this pressure, the resistance does not exhibit a strong anisotropy, thus
the ground state past 10~kbar is a rotationally invariant, uniform electron fluid. 
At $\nu=5/2$ we find a second quantum phase transition near $P \simeq 10$~bar
between the nematic phase and an isotropic Fermi liquid.
We note that in Fig.2 we also see weak FQHSs at $\nu=7/3$ and $8/3$. However, the nematic phase
is not stabilized at these filling factors.

We now discuss possible origins for the observed isotropic FQHS-to-nematic phase transition. Pioneering
numerical work found that  a transition from the Pfaffian to a nematic phase occurs
when the effective electron-electron interaction is tuned away from its Coulomb expression \cite{haldane2000}.
In this work the required interaction was generated by varying the layer width $w$ of the 2DEG. 
The layer width appears in the theory in its adimensional form $w/l_B$, 
where $l_B=\sqrt{\hbar/eB}$ is the magnetic length.
Later theoretical work addressed the evolution of the stability of the Pfaffian at $\nu=5/2$
with $w/\l_B$ \cite{thickness1,thickness2} and with the magnitude of a three-body interaction  
\cite{wan,wang}. However, the lack of a spontaneously developed nematic phase at $\nu=5/2$ 
in experiments at ambient pressures suggests that tuning of the
layer width by itself is not sufficient to stabilize the nematic phase.

Landau level mixing, as measured by the Landau level mixing parameter $\kappa$, 
also affects the effective electron-electron interaction \cite{yoshi}. Theory of the ground state at $\nu=5/2$
in the $\kappa- w/l_B$ plane, however, has not considered an instability toward a nematic
phase \cite{llm1,llm2,llm3,param,yuli}. We think
that by tuning the pressure, we access a combination of $\kappa$ and $w/l_B$, which in the
spirit of Ref.\cite{haldane2000}, stabilizes the nematic phase. A detailed discussion of the dependence on pressure
of these parameters can be found in the Supplement. Decreasing density with an increasing pressure
\cite{densP} has by far the strongest imprint on these values. We find that 
in the pressure range $P=8.26 \div 9.76$~kbar at which the nematic phase is detected, 
the density ranges between $n = 1.0 \div 0.68 \times 10^{11}$/cm$^2$, the Landau level mixing parameter
spans $\kappa = 2.08 \div 2.57$, and the adimensional layer width spreads over $w/l_B = 1.51 \div 1.24$.
Simply put, the nematic phase in our experiment develops when the conditions $\kappa>2$ and $w/l_B<1.5$ are
simultaneously met. From a survey of the literature on electron samples we find that the above region of 
the $\kappa$-$w/l_B$ parameter space has not yet been accessed \cite{pan14,nodar,pan12,pan15}. 
We thus think that either the combination of low enough $w/l_B$ and high enough $\kappa$
or other yet unknown factors are responsible for the appropriate
short-range interactions needed for stabilizing the nematic phase at $\nu=5/2$.
Nonetheless, clarifying the stability conditions of the nematic phase at $\nu=5/2$
needs further investigation since this phase does not develop
in $p$-doped samples with $\kappa=14.4$ and $w/l_B=1.1$ \cite{manfra}
or  $\kappa=20$ and $w/l_B=0.79$ \cite{kumar}.

Finally, we note that the nematic phase we referred to is of electrons. We notice that 
at $P = 8.26$~kbar there is a weak FQHS present at $\nu=8/3$ in the vicinity of the nematic phase.
The presence of this weak FQHS signals the formation of composite fermions, thus
there is an interesting possibility that anisotropy observed in Fig.1c is due 
to a nematic phase of composite fermions \cite{SYLee01}.

We measured a symmetrically doped 
$w=30$~nm wide Al$_{0.24}$Ga$_{0.76}$As/GaAs/Al$_{0.24}$Ga$_{0.76}$As quantum well
sample \cite{deng,manf}. The density measured at ambient pressures is  $n=2.8 \times 10^{11}$cm$^{-2}$ and the mobility is
$15 \times 10^6$cm$^2$/Vs. The high hydrostatic pressures were generated using a commercial
pressure clamp cell \cite{cell}. Each incremental increase in the pressure was done at room temperature;
pressures were measured both at room temperature as well as at low temperatures using different manometers.
Pressures quoted throughout our paper, however, are the low temperature pressures. The pressure cell also
contained a small light emitting diode which was used
to illuminate the sample after every cooldown in order to prepare the state.
We chose a sample size of $2 \times 2$ mm$^2$ lateral extent so that it can easily fit inside the teflon lining
of the pressure cell. As shown in Fig.1, our square shaped sample has
four indium ohmic contacts on the corners of the square. The longitudinal resistance was measured using
a standard low frequency lockin technique at an ac excitation of 2~nA.

Sample growth and measurement at Purdue were supported by the US Department of Energy, Office of Basic Energy Sciences,
Division of Materials Sciences and Engineering under the award DE-SC0006671. E.F. acknowledges the US National Science Foundation grant DMR 1408713. We thank J.P. Eisenstein for his comments and Dr. Marcio Siqueira for advice on using the pressure cell.


\begin{thebibliography}{l}

\bibitem{fogler} Koulakov, A.A., Fogler, M.M. \& Shlovskii, B.I. Charge density wave in two-dimensional
electron liquid in weak magnetic field. \textit{Phys. Rev. Lett}. \textbf{76}, 499-502 (1996).
\bibitem{moessner} Moesnner, R. \& Chalker, J.T. Exact results for interacting electrons in high Landau levels.
\textit{Phys. Rev. B} \textbf{54}, 5006-5015 (1996).

\bibitem{kivelson} Fradkin, E., Kivelson, S.A., Lawler, M.J., Eisenstein, J.P. \& Mackenzie, A.P.
Nematic Fermi fluids in condensed matter physics. \textit{Annu. Rev. Conden. Matter Phys.}
\textbf{1}, 153-178 (2010).
\bibitem{fradkin} Fradkin, E. \& Kivelson S.A. Liquid-crystal phases of quantum Hall systems. \textit{Phys. Rev. B}
\textbf{59}, 8065-8072 (1999).
\bibitem{lilly} Lilly, M.P., Cooper, K.B., Eisenstein, J.P., Pfeiffer, L.N. \& West, K.W. Evidence for
an anisotropic state of two-dimensional electrons in high Landau levels. \textit{Phys. Rev. Lett}. \textbf{82},
394-397 (1999).
\bibitem{du} Du, R.R., Tsui, D.C., Stormer, H.L., Pfeiffer, L.N., Baldwin, K.W. \& West, K.W. Strongly
anisotropic transport in higher two-dimensional Landau levels. \textit{Solid State Commun.} \textbf{109},
389-394 (1999).
\bibitem{heterostructure} Pollanen, J., Cooper, K. B., Brandsen, S., Eisenstein, J. P., Pfeiffer, L. N., \& West, K. W. Heterostructure symmetry and the orientation of the quantum Hall nematic phases. \textit{Phys. Rev. B.} \textbf{92}, 115410 (2015).

\bibitem{high1}  Tranquada, J.,  Sternlieb, B.J., Axe, J.D., Nakamura, Y. \& Uchida, A. Evidence for stripe correlations of spins and holes in copper oxide superconductors.  \textit{Nature} \textbf{375}, 561-563 (1995).
\bibitem{high3} da Silva Neto, E.H., Aynajian, P., Frano, A., Comin, R., Schierle, E., Weschke, E., Gyenis, A., Wen, J., Schneeloch, J., Xu, Z., Ono, S., Gu, G., Le Tacon, M. \& Yazdani, A. Ubiquitous interplay between charge ordering and high-temperature superconductivity in cuprates. \textit{Science} \textbf{343}, 393-396 (2014).
\bibitem{high4} Fradkin, E., Kivelson, S.A.  \& Tranquada, J.M. \textit{Colloquium:} Theory of intertwined orders in high temperature superconductors.  \textit{Rev. Mod. Phys.}  \textbf{87}, 457 (2015).
\bibitem{he3} Vortontsov, A.G.  \& Sauls, J.A. Crystalline order in superfluid $ ^{3}\mathrm{He}$ films.  \textit{Phys. Rev. Lett.}  \textbf{98},  045301 (2007).

\bibitem{landau} Landau, L.D. \& Lifshitz, E.M. \textit{Statistical Physics, Course of theoretical physics, Vol. 5, 3rd ed.} (Butterworth-Heinemann, Oxford, UK, 1980).
\bibitem{wen1} Wen, X.G. \textit{Quantum field theory of many-body systems.} (Oxford University Press, Oxford, UK, 2004).
\bibitem{klitzing} von Klitzing, K., Dorda, G. \& Pepper, M. New method for high-accuracy determination of the fine-structure constant based on quantized Hall resistance. \textit{Phys. Rev. Lett}. \textbf{45}, 494-497 (1980).
\bibitem{tsui} Tsui, D.C., Stormer, H.L. \& Gossard, A.C. Two-dimensional magnetotransport in the extreme quantum limit. \textit{Phys. Rev. Lett}. \textbf{48}, 1559-1562. (1982).	

\bibitem{read} Moore, G. \& Read, N. Nonabelions in the fractional quantum Hall effect. \textit{Nucl. Phys. B}. \textbf{360}, 362-396 (1991).
\bibitem{willett} Willett, R.,  Eisenstein, J.P., Stormer, H.L., Tsui, D.C., Gossard, A.C. \&  English J.H.
Observation of an even-denominator quantum number in the fractional quantum Hall effect. 
\textit{Phys. Rev. Lett}. \textbf{59}, 1776-1779 (1987).
\bibitem{panq} Pan, W., Xia, J.-S., Shvarts, V., Adams, D.E., Stormer, H.L., Tsui, D.C., Pfeiffer, L.N.,
Baldwin, K.W. \& West, K.W. Exact quantization of the even-denominator fractional quantum 
Hall state at $\nu=5/2$ Landau level filling factor. \textit{Phys. Rev. Lett.} \textbf{83}, 3530-3533 (1999).

\bibitem{haldane2000} Rezayi, E.H. \& Haldane, F.D.M. Incompressible paired Hall state, stripe order, and the composite fermion liquid phase in the half-filled Landau levels. \textit{Phys. Rev. Lett.} \textbf{84}, 4685-4688 (2000).
\bibitem{wan} Wan, X., Hu, Z. X., Rezayi, E. H. \& Yang, K. Fractional quantum Hall effect at $\nu=5/2$: ground states, non-Abelian
quasiholes, and edge modes in a microscopic model. \textit{Phys. Rev. B} \textbf{77}, 165316 (2008).
\bibitem{wang} Wang, H., Sheng, D.N. \& Haldane, F.D.M. Particle-hole symmetry breaking and the $\nu=5/2$ fractional
quantum Hall effect. \textit{Phys. Rev. B} \textbf{80}, 241311 (2009).

\bibitem{pan99}  Pan, W., Du, R.R., Stormer, H.L., Tsui, D.C., Pfeiffer, L.N., Baldwin, K.W. \& West, K.W.
Strongly anisotropic electronic transport at Landau level filling factor $\nu=9/2$ and $\nu=5/2$ under tilted magnetic field.
\textit{Phys. Rev. Lett}. \textbf{83}, 820-823 (1999).
\bibitem{lilly99} Lilly, M.P., Cooper, K.B., Eisenstein, J.P., Pfeiffer, L.N. \& West, K.W. Anisotropic states of two-dimensional 
electron systems in high Landau levels: effect of an in-plane magnetic field. \textit{Phys. Rev. Lett}. \textbf{83},
824-827 (1999).

\bibitem{friess14} Friess, B., Umansky, V., Tiemann, L., von Klitzing, K. \& Smet, J.H.
Probing the microscopic structure of stripe phase at filling factor $5/2$. \textit{Phys. Rev. Lett}. \textbf{113}, 076803 (2014).
\bibitem{pan15} Shi, X., Pan, W., Baldwin, K.W., West, K.W., Pfeiffer, L.N.
\& Tsui, D.C. Impact of the modulation doping layer an the $\nu=5/2$ anisotropy.
\textit{Phys. Rev. B} \textbf{91}, 125308 (2015).
\bibitem{xia10} Xia, J., Cvicek, V., Eisenstein, J.P., Pfeiffer, L.N. \& West, K.W. Tilt induced anisotropic 
to isotropic phase transition at $\nu=5/2$. \textit{Phys. Rev. Lett}. \textbf{105}, 176807 (2010).
\bibitem{xia11} Xia, J., Eisenstein, J.P., Pfeiffer, L.N. \& West, K.W. Evidence for a fractionally 
quantized Hall state with anisotropic longitudinal transport. \textit{Nature Phys.} \textbf{7}, 845-848 (2011).
\bibitem{liu13} Liu, Y., Hasdemir, S., Shayegan, M., Pfeiffer, L.N., West, K.W. \& Baldwin, K.W.
Evidence for a $\nu=5/2$ fractional quantum Hall nematic state in parallel magnetic fields.
\textit{Phys. Rev. B}. \textbf{88}, 035307 (2013).



                          
\bibitem{qh} Prange, R.E. \& Girvin, S.M. \textit{The quantum Hall effect.} (Springer-Verlag, New York, 1987).
                                            
\bibitem{pan14} Pan, W., Serafin, A., Xia, J. S., Yin, L., Sullivan, N.S., Baldwin, K W., West, K.W., Pfeiffer, L.N.
\& Tsui, D.C. Competing quantum Hall phases in the second Landau level in the low-density limit.
\textit{Phys. Rev. B} \textbf{89}, 241302 (2014).

\bibitem{leo11} Koduvayur, S.P., Lyanda-Geller, Y., Khlebnikov, S., Cs\'athy, G.A., Manfra, M.J., Pfeiffer, L.N.,
West, K.W. \& Rokhinson, L.P. Effect of strain on stripe phases in the quantum Hall regime. \textit{Phys. Rev. Lett}.
\textbf{106}, 016804 (2011).

\bibitem{ran} Ran, Y. \& Wen, X.G. Detecting topological order through a continuous quantum phase transition.
\textit{Phys. Rev. Lett}. \textbf{96}, 026802 (2006).
\bibitem{fr2} Ardone, E., Fendley, P. \& Fradkin, E. Topological order and conformal quantum critical points. \textit{Ann. Phys.} \textbf{310}, 493 (2004).




\bibitem{nodar} Samkharadze, N., Watson, D.J., Gardner, G. Manfra, M.J., Pfeiffer, L.N., West, K.W. \& Cs\'athy,
G.A. Quantitative analysis of the disorder broadening and the intrinsic gap for
the $\nu=5/2$ fractional quantum Hall state. \textit{Phys. Rev. B} \textbf{84}, 121305 (2011).
\bibitem{pan12} Pan, W., Baldwin, K.W., West K.W., Pfeiffer, L.N. and Tsui, D.C.
Spin transition in the $\nu=8/3$ fractional quantum Hall effect.  \textit{Phys. Rev. Lett.}
\textbf{108}, 216804 (2012).

\bibitem{thickness1} Peterson, M.R., Jolicoeur, Th. \& S. Das Sarma, S.
Finite-layer thickness stabilizes the Pfaffian state for the 5/2 fractional quantum Hall effect: 
wave function overlap and topological degeneracy. \textit{Phys. Rev. Lett.} \textbf{101}, 016807 (2008).
\bibitem{thickness2} Papi\'c, Z., Regnault, N. \& Das Sarma, S. Interaction-tuned compressible-to-incompressible 
phase transitions in quantum Hall systems. \textit{ Phys. Rev. B} \textbf{80}, 201303 (2009).
\bibitem{yoshi} Yoshioka, D. Excitation energies of the fractional quantum Hall effect. \textit{J. Phys. Soc. Jpn.}, \textbf{55}, 885-896 (1986).
\bibitem{llm1} W\'ojs, A. \& Quinn, J.J. Landau level mixing in the $\nu=5/2$ fractional quantum Hall state. \textit{Phys. Rev. B} \textbf{74}, 235319 (2006).
\bibitem{llm2} W\'ojs, A., T\H oke, C. \& Jain, J.K. Landau-level mixing and the emergence of Pfaffian excitations for the 5/2 fractional quantum Hall effect. \textit{Phys. Rev. Lett.} \textbf{105}, 096802 (2010).
\bibitem{llm3} Nuebler, J., Umansky, V., Morf, R., Heiblum, M., von Klitzing, K. \& Smet, J.
Density dependence of the $\nu=5/2$ energy gap: Experiment and theory. \textit{Phys. Rev. B} \textbf{81}, 035316 (2010).
\bibitem{param} Pakrouski, K., Peterson, M.J. Jolicoeur, Th., Scarola, V.W., Nayak, C. \& Troyer, M.
Phase diagram of the $\nu=5/2$ fractional quantum Hall effect: effects
of Landau-level mixing and nonzero width. \textit{Phys. Rev. X} \textbf{5}, 021004 (2015).
\bibitem{yuli} Tylan-Tyler, A. \& Lyanda-Geller, Y.  Phase diagram and edge states of the $\nu=5/2$ fractional 
quantum Hall state with Landau level mixing and finite well thickness. \textit{Phys. Rev. B} \textbf{91}, 205404 (2015).

\bibitem{densP} Dmowski, L. \& Portal, J.C. Magnetotransport in 2D semiconductor systems under pressure.
\textit{Semicond. Sci. Technol.} \textbf{4}, 211-217 (1989). 

\bibitem{manfra} Manfra, M.J., de Piciotto, R., Jiang, Z.,  Simon, S.H., Pfeiffer, L.N., West, K.W. \& Sergent, A.M.
Impact of spin-orbit coupling on the quantum Hall nematic phases. \textit{Phys. Rev. Lett.} \textbf{98}, 206804 (2007).
\bibitem{kumar} Kumar, A., Samkharadze, N., Cs\'athy, G.A., Manfra, M.J., Pfeiffer, L.N. and West K.W.
Particle-hole asymmetry of fractional quantum Hall states in the second Landau level of a two-dimensional hole system.
\textit{Phys. Rev. B} \textbf{83}, 201305 (2011).

\bibitem{SYLee01} Lee, S.-Y., Scarola, V.W. \& Jain, J.K. Stripe formation in the fractional quantum Hall regime. \textit{Phys. Rev. Lett}. \textbf{87}, 256803 (2001).

\bibitem{deng} Deng, N., Watson, J.D., Rokhinson, L.P., Manfra, M.J. \& Cs\'athy, G.A.
Contrasting energy scales of reentrant integer quantum Hall states. \textit{Phys. Rev. B} \textbf{86}, 201301 (2012).
\bibitem{manf} Manfra, M.J. Molecular beam epitaxy of ultra-high-quality AlGaAs/GaAs heterostructures:
enabling physics in low-dimensional electronic systems. \textit{Annu. Rev. Condens. Matter Phys.} \textbf{5}, 347-373 (2014).
\bibitem{cell} easyLab Technologies Ltd, model Pcell 30.


\end{thebibliography}
\end{document}